# The strongest size in the inverse Hall-Petch relationship

Yong Pan,[1] Yichun Zhou,[1,♥] Chang Q Sun[2*]

[1] Key laboratory of low-dimensional materials and application technologies (Xiangtan University), Ministry of Education, Hunan 411105

[2] School of Electrical and Electronic Engineering, Nanyang Technological University, Singapore 639798

Abstract

Incorporating the bond-order-length-strength correlation mechanism [Sun CQ, *Prog Solid State Chem* 35, 1 -159 (2007)] and Born's criterion for melting [*J. Chem. Phys.* 7, 591(1939)] into the conventional Hall-Petch relationship has turned out an analytical expression for the size and temperature dependence of the mechanical strength of nanograins, known as the inverse Hall-Petch relationship (IHPR), that has long been a topic under debate regarding the possible mechanisms. Reproduction of the measured IHPR of Ni, NiP and $TiO_2$ nanocrystals revealed that: (i) the size induced energy densification and cohesive energy loss of nanograins originates the IHPR that could be activated in the contact mode of plastic deformation detection; (ii) the competition between the inhibition of atomic dislocations, via the surface energy density gain and the strain work hardening, and the activation for dislocations through cohesive energy loss determine the entire IHPR profile of a specimen; (iii) the presence of a soft quasisolid phase is responsible for the size-induced softening and the superplasticity as well of nanostructures; (iv) the bond nature involved and the $T/T_m$ ratio between the temperature of operating and the temperature of melting dictate the measured strongest sizes of a given specimen.

[♥] zhouyc@xtu.edu.cn
[*] Associated at Xiantan University, ecqsun@ntu.edu.sg

**I  Introduction**

The mechanically strengthening with grain refinement in the size range of sub-micrometers has traditionally been rationalized with the conventional temperature-independent Hall-Petch relationship (HPR)[1] that can be simplified in a dimensionless form being normalized by the bulk standard, P(0), measured under the same conditions,

$$P(x)/P(0) = 1 + Ax \qquad (1)$$

the slope $A$ is an adjustable parameter used to fit experimental data. The K = R/d, and x = $K^{-1/2}$, is the dimensionless form of size, which corresponds to the number of atoms, with mean diameter d, lined along the radius, R, of a spherical-like nanograin, as illustrated in Figure 1. Using the dimensionless form of the mechanical strength and the grain size aims to minimize the contribution from artifacts such as the processing conditions and the crystal orientation. The HPR has been well understood in terms of the pile up of dislocations that resist plastic flow from grain refinement.[2,3]

However, as the crystal is refined from the micrometer regime into the nanometer scale, the HPR process invariably breaks down and the relationship of yield strength versus grain size departs markedly from that seen at larger grain sizes. With further grain refinement, the yield strength peaks, in many cases, at a mean grain size in the order of 10 nm or so. A further decrease in grain size can cause softening of the solid, instead, and then the HPR slope turns from positive to negative at a critical size, or called the strongest size. The deviation from the HPR is called the inverse Hall-Petch relationship (IHPR).[4,5,6,7,8,9]

There is a concerted global effort underway towards deeper insight into the IHPR mechanism with postulated explanations in terms of dislocation-based,[10] diffusion-based,[11,12] grain-boundary-shearing-based,[13] core-shell-role-exchange-based,[14] two-phase-based,[15] collective-dislocation-based,[16] and dislocation-absorption-based[17] models. It has been suggested that the grain boundaries (GBs) consisting of under-coordinated atoms contribute to the GB performance.[18] The strongest Cu grain size of 10 ~ 15 nm, for instance, is attributed to a switch in the microscopic deformation mechanism from dislocation-mediated plasticity in the coarse-grain interior to the GB sliding in the nanocrystalline regime.[19] A significant portion of atoms resides in the GB and the plastic flow of the GB region is responsible for the unique characteristics displayed by such materials.[20] In the HPR regime, crystallographic slips in the grain interiors govern the plastic behavior of the polycrystallite; while in the IHPR regime, GB dominates the plastic behavior. During the transition, both grain interior and GB contribute competitively. The slope in the HPR is suggested to be proportional to the work required to eject dislocations from GBs.[21] The strongest size is suggested to depend strongly on the stacking-fault energy and the magnitude of the applied stress.[22,23]

Although there is a growing body of experimental evidence for such unusual deformation in the nanometer regime, the underlying atomistic mechanisms for the IHPR are yet poorly understood. As pointed out by Kumar et al[24] and Mayrhofer et al [25] the physical origin of the IHPR transition and the factors dominating the strongest size has been a long-standing puzzle. The objective of this contribution is to show that a combination of the conventional HPR,[1] Born's criterion for melting,[26] and the recently developed bond-order-length-strength (BOLS) correlation mechanism[27] for the size dependence of nanostructures has allowed us to derive an analytical a solution, which enable us to clarify the factors dictating the strongest size and the physical origin of the observed IHPR.

## II Principle

### 2.1 BOLS correlation and Bern's criterion

According to Born,[26] the modulus of a specimen attenuates when the operating temperature is raised and the modulus approaches to zero at the melting point ($T_m$). Therefore, the $T_m$ is a key character to the mechanical strength of a material, in particular, for the plastic deformation detection. It is understandable that a liquid phase is extremely soft and highly compressible when it is being pressed or stretched. Hence, the separation between the $T_m$ and the temperature of operation is crucial to the mechanical behavior of a material. If one could lower the $T_m$ by reducing the solid size,[27] the materials will become softer and softer as the solid size become smaller and smaller under the same operating temperature.

The BOLS correlation mechanism suggests that:[27]

(i) If one bond breaks, the neighboring ones become shorter and stiffer. Hence, local strain and energy trapping take place at sites surrounding defects or at the surface skin of a substance disregarding the surface curvature. This effect is associated with local densification of charge, energy, and mass in the surface skin that serves then as pinning center inhibiting atomic dislocations.

(ii) On the other hand, the broken bonds will lower the atomic cohesive energy of the under-coordinated atoms (the product of bond number and the bond energy), which dictates the critical temperatures for phase transition such as melting and evaporating. The cohesive energy also determines the activation energy for atomic dislocation or diffusion. Therefore, defects also provide sites initiating structure failure.

Figure 1 illustrates the core-shell structure of a nanograin of K radius with strained surface skin of 2~3 atomic diameter thick. The under-coordinated atoms in the skin are the key to nanostructures yet atoms in the core interior remain their bulk nature. Because of the broken surface bonds and the broken bond induced local strain and trapping, the energy density is higher and the atomic cohesive energy is lower in the surface skin in comparison to that in the core interior. The surface shell is harder at low temperatures because of the localised high energy density yet the skin melts more easily because of the lowered atomic cohesive energy. When the temperature approaches the surface meting, the surface atoms dislocate more easily and hence the skin is much softer than the grain interior. As the grain size is decreased the skin effect will be more pronounced because of the increased portion of the under-coordinated atoms.

2.2  Relative stress and modulus at surface

Intrinsically, the stress, flow stress or hardness, and the elastic modulus being the same in dimension and the same unit of Pa, or $J/m^3$, are proportional to the sum of bond energy per unit volume.[28] However, in the contact mode of plastic deformation detection, processes of activation and inhibition of atomic dislocations are involved simultaneously, which make the hardness differs from the elastic modulus in observations.

For a given specimen, the nature and the total number of bonds remain unchanged unless phase transition occurs. However, the length and strength of the involved bonds vary with external stimulus such as the coordination environment and the operating temperature. Therefore, we need to consider only the binding energy of a representative bond and the volume of an atom in deriving the T-dependent stress and bulk modulus at the site of the ith atom with $z_i$ coordinates:[29]

$$P_i(z_i,T) = -\left.\frac{\partial u(r,T)}{\partial V}\right|_{d_i,T} \propto B_i = V\left.\frac{\partial P_i}{\partial V}\right|_{d_i,T} \propto Y_i \sim \begin{cases} \propto \dfrac{E_i(0)}{d_i^3} & (T \ll T_{m,i}) \\ = \dfrac{E_{1i}(T)}{d_i^3(1+\alpha T)} & (else) \end{cases}$$

$$E_i(T) = E_i(0) - \int_0^T \eta_{1i}(t)dt \stackrel{T \gg \theta_D}{=} \eta_{2i} + \eta_{1i}(T_{m,i} - T)$$

(2)

where u(r) is a pairing potential, $E_i$ is the bond energy. The linear thermal expansion coefficient is in the range of $\alpha = 2\sim4\times10^{-7}/T_m$, which can be neglected in numerical calculations compared to the extent of size-induced bond contraction that is in the order of ~1-10%. The $\eta_{1i}$ is the specific heat per bond. $\theta_D$ is the Debye temperature. The $\eta_{2i}$ is $1/z_i$ fold of the latent heat of atomization for the ith atom from molten state. If Bern's criterion applies, the $\eta_{2i}$ can be omitted in the analytical expression. Therefore, the normalized $P_i$ and $B_i$ by the corresponding bulk values, denoted with subscript b, share the commonly dimensionless form:

$$\frac{P_i(z_i,T)}{P(z_b,T_0)} = \frac{B_i(z_i,T)}{B(z_b,T_0)} \begin{cases} \propto c_i^{-(m+3)} & (T \ll T_{m,i}) \\ = \dfrac{\eta_{1i}d^3(T_{m,i}-T)}{\eta_{1b}d_i^3(T_{m,b}-T)} = \dfrac{\eta_{1i}c_i^3(T_{m,i}/T_{m,b}-T/T_{m,b})}{\eta_{1b}(1-T/T_{m,b})} & (else) \end{cases}$$

with,[27]

$$c_i(z_i) = d_i/d = 2/(1+\exp((12-z_i)/(8z_i)))$$
$$c_i^{-m} = E_i(0)/E_b(0)$$
$$T_{m,i} \propto z_i/z_b T_{m,b} = z_{ib}T_{m,b}$$
$$z_i = 4(1-0.75/K); z_2 = z_1 + 2; z_{i\leq3} = z_b = 12.$$

(3)

where $c_i(z_i)$ is the bond contraction coefficient and m the bond nature indicator. For elemental metals, m = 1 unless surface contamination occurs; for alloys or compounds, the m ~ 4.[27] The $\eta_{1i}/\eta_{1b}$ value of 3.37 has been obtained for an impurity-free gold monatomic chain.[30] The $B_i$ and $P_i$ depends uniquely on the bond length and bond energy that vary with the coordination environment, $(T_{m,i}, c_i)$, bond nature, (m), and the relative temperature $(T/T_{m,b})$.

2.3 IHPR: strongest size and quasi-solid state

By substituting the size and bond nature dependent $\eta_1(x)$, $d(x)$, and $T_m(x, m)$ for the $\eta_{1i}$, $d_i$, and $T_{m,i}$ in eq (3), we can readily obtain the size, bond nature, and temperature dependence of the mechanical strength of a nansolid,

$$\frac{P(x,T)}{P(0,T)} = \frac{\eta_1(x)}{\eta_1(0)}\left(\frac{d(0)}{d(x,T)}\right)^3 \times \frac{T_m(x,m)-T}{T_m(0)-T} = \frac{P(x)}{P(0)}\varphi(d,x,m,T)$$

(4)

The additional term $\varphi(d, x, m, T)$ to the conventional HPR, $P(x)/P(0)$, covers contributions from the bond nature, temperature, and the size and bond nature dependent $d(x)$ and $T_m(x, m)$ of a nanocrystal, which are in the form of:[27]

$$\begin{cases} d(x)/d(0) = 1 + \Delta_d(x); & \Delta_d(x) = \sum_{i\leq 3}\gamma_i(x)\times(c_i-1) \\ T_m(x,m)/T_m(0) = 1 + \Delta_B(x,m); & \Delta_B(x,m) = \sum_{i\leq 3}\gamma_i(x)\times(z_{ib}c_i^{-m}-1) \\ \gamma_i(x) = \tau c_i K^{-1} = \tau c_i x^2 & (K \geq \tau) \end{cases}$$

(5)

the $\Delta_d(x)$ and $\Delta_B(x, m)$ are the broken bonds induced perturbations to the bond length and to the atomic cohesive energy, $E_B$. $\gamma_i(x)$ is the volume or number ratio of atoms in the ith atomic layer to that of the entire solid. The parameter $\tau$ is the dimensionality for spherical dot ($\tau = 3$), cylindrical rod ($\tau = 2$), and thin plate ($\tau = 1$). The sum is carried out over the outermost three atomic layers.

By comparing the currently derived form, eq (**4**), with the traditionally T-independent HPR, eq (1), one can readily find the relation, $\eta_1(x)/\eta_1(0) = P(x)/P(0) = 1 + Ax$, which differs the hardness from the elastic modulus.

This term should represent the HPR in which the resistance and activation of atomic dislocations contribute competitively.

Incorporating the activation energy for atomic dislocations, $E_A \propto T_m$ [Refs 31,32,33], into the prefector $A$, leads to the analytical expression for the size, bond nature, and temperature dependent HPR:

$$\begin{aligned}
\frac{P(x,T)}{P(0,T)} &= \frac{\eta_1(x)}{\eta_1(0)} \left(\frac{d(0)}{d(x)}\right)^3 \times \frac{T_m(x,m) - T}{T_m(0) - T} \\
&= \frac{\eta_1(x)}{\eta_1(0)} [1 + \Delta_d(x)]^{-3} \times \frac{T_m(0)[1 + \Delta_B(x,m)] - T}{T_m(0) - T} \\
&= \frac{1 + A(x, \theta(x))}{[1 + \Delta_d(x)]^3} \times \left[1 + \frac{\Delta_B(x,m)}{1 - \theta(T)}\right]
\end{aligned}$$

(6)

with

$$A(x, \theta(x)) = A_0 x \exp\left[\frac{T_m(x)}{T}\right] = A_0 x \exp\left[\frac{1 + \Delta_B(x,m)}{\theta(T)}\right]$$

(6b)

where $\theta(T) = T/T_m(0)$. The prefactor $A_0$ is an adjustable parameter. The expression indicates that:

(i) The IHPR originates from the size-induced change of the mean bond length ($\Delta_d(x)$) and the $T_m(\Delta_B(x))$ of the specimen. At sufficiently large sizes, the $\Delta_B(x,m)$ and the $\Delta_d(x)$ approach to zero and the derived form converges to the conventional HPR.

(ii) The competition between the cohesive energy loss, $\Delta_B(x,m) < 0$, and the energy density gain, $\eta_1(x)/d^3(x) \gg 1$, of the nanograins and the competition between the activation ($E_A \propto T_m(x)$ drops with K) and the inhibition (x or $K^{-1/2}$ increases) dictate the IHPR observations.

(iii) It is further clarified that the parameter A in the HPR is replaced by $A_0 \exp(T_m(0)/T)$ and the classical HPR becomes $P/P_0 = 1 + A_0 x \times \exp(T_m(0)/T)$, representing the competition between activation and inhibition of atomic dislocations.

According to this solution, the GB is harder at temperatures far below the $T_m(x, m)$ because of the dominance of energy density gain, whereas at temperatures close to $T_m(x, m)$, the GB is softer than the grain interior because of the atomic cohesive energy loss that lowers the barrier for atomic dislocation. Hence, the current form supports for all the explanations given by the available models.[10-24]

The strongest size, $x_C(f, \theta(T), m)$, and the dominating factors for the strongest size, can be determined by the relation, $d(Ln(P/P_0))/d(\ln x) = 0$,

$$\frac{d(Ln(P/P_0))}{d(\ln x)} = x \frac{d(Ln(P/P_0))}{dx} = A(x, \theta(x), m) \times \frac{\theta(0) + 2\Delta_B(x,m)}{\theta(0)(1 + A(x, \theta(x), m))}$$

$$-\frac{6\Delta_d(x)}{1+\Delta_d(x)} + \frac{2\Delta_B(x,m)}{1+\Delta_B(x,m)-\theta(0)} = 0$$

or,

$$A(x, \theta(x), m) \times \frac{1 + 2\Delta_B(x,m)/\theta(0)}{1 + A(x, \theta(x), m)} - \frac{6}{1+\Delta_d(x)} + \frac{2(\theta(0)-1)}{1+\Delta_B(x,m)-\theta(0)} = 4$$

(7)

where P represents for the $P(x, \theta(x))$ and $P_0$ for the $P(0, \theta(0))$.

As discussed above, a nanograin becomes softer with respect to the bulk if the grain is sufficient small. It is necessary to define the critical temperature, $T_C$, for the solid-quasisolid and quasisolid-liquid transition as follows:

$$\frac{P(x, T_C)}{P(0, T_C)} = \frac{1 + A(x, \theta(T_C), m)}{[1+\Delta_d(x)]^3} \times \left[1 + \frac{\Delta_B(x,m)}{1-\theta(0)}\right] = \begin{cases} 1 & (Quasisolid) \\ 0 & (Liquid) \end{cases}$$

(8)

At temperature higher than the $T_C(x, A_0, m)$, the solid is softer and easily compressible compared with the bulk counterpart at the same temperature. At the $T_m(x, m)$ or higher, the nanograin becomes liquid that is extremely soft and highly compressible, complying with Born's criterion for melting. As it will be shown shortly, the $T_C$ for solid-quasisolid transition is size dependent and it is much lower than the $T_m$.

III    Results and discussion

3.1    Characteristics of nanograins

Figure 2 shows the BOLS reproduction of the size-induced energy density gain represented by the modulus enhancement of ZnO nanowires,[34] and the cohesive energy loss represented by the melting point depression of Al[35] and Sn[36] plates and the evaporating temperature depression of Au and Ag dots[37]. Prediction also agrees with the size trend of Young's modulus enhancement for Ag nanowires[38] and the measured surface inward relaxation and the mean lattice contraction of nanostructures.[27] These observations evidence the significance of the under-coordinated atoms in the surface skins of nanostructures to the size induced property change. The broken bonds seemed contribute not directly to observations but the consequence of bond breaking is indeed profound and significant. Interested readers are referred to Ref 27 for more details regarding the size effect on the thermal and chemical stability, lattice dynamics, optical emission and absorption, electronic, dielectric and magnetic behaviour of nanostructures that follow consistently the BOLS predictions.

3.2    Phase diagram: quasisolid state

Eq (**8**) gives rise to the $T_C$-K phase diagram contains the outstanding regions of solid, quasisolid, and liquid states, as shown in Figure 3a. The two $T_m$-K profiles being overlapped are derived from the BOLS correlation (eq (**5**)) and the Bern's criterion for melting (eq (**8**)). For a given size, the $T_C$ is much lower than the $T_m$. The $T_C$ drops even faster than the $T_m$ when the solid size is reduced. For a Cu nanosolid with K = 10 (~5 nm in diameter) instance, the bond contracts by a mean value of 5%, associated with a 25% drop of $T_m$ and a 50% drop of $T_C$ with respect to the bulk $T_m(0)$ (1358 K). The 5 nm-sized Cu being in a quasisolid state at 680 K or above will be softer than the bulk counterpart at the same temperature. This understanding may explain why the strength of 300-nm-sized Cu nanograins drops by 15% associated with substantial increase of the ductility measured at 500 K.[39] On the other hand, the self-heating in operation should raise the actual temperature and hence cause further softening of the specimen. In the contacting mode of plastic deformation testing, the bond breaking and deforming will releases energy that will heat up the specimen considerably. Hence, the size-induced $T_C$ drop and the self-heating in operation provide mechanism for the softening in the IHPR and the high ductility of metallic nanowires,[40,41] as well.

Figure 3b shows temperature dependence of the relative hardness of metallic particles of two critical sizes. The hardness drops quickly when the temperature is raised. At the quasisolid state, the $P/P_0$ is smaller than unity. At the $T_m$, the $P/P_0$ approaches to zero. Compared with the nonlinear $P/P_0$ –T relationship, the $Y/Y_0$ depends linearly on T [ref 42] at $T > \theta_D$ because of the absence of the competitive processes of activation and inhibition of atomic dislocations in the hardness measurement.

3.3   IHPR and the strongest size

Calculations using eqs (**5**) and (**6**) were performed on Ni[43], NiP[44], and TiO$_2$[45] nanograins with the standard bulk d and $T_m$ values and T = 300 K as input. The prefector *A₀* is adjusted under the constraint that the slope of the traditional HPR (straight lines) should match to observations and the curve should intercept at the positive side of the vertical axis. For comparison purpose, the theoretical curves were normalized with the calculated peak values at $x_C$, and the experimental data measured at room temperature were normalized with the measured peak values. Figure 4 shows the reproduced IHPR for the given specimens. The solid lines are the current IHPR using the same $A_0$ value for the corresponding HPR. The dashed lines only consider the competition between the activation and inhibition of dislocations with the term $\varphi(d, m, x, T) = 1$ in eq **(4)**, being quite the same to the approach of Zhao et al.[31] Reproduction of the measured data evidence the validity of the IHPR that is dominated by the competition between the activation and the inhibition of atomic dislocations represented by $(1+A_0 x \times \exp(T_m(x)/T))$ though the term $\varphi(d, T, m)$ is necessary.

Based on eqs (5) and (7) with the atomic diameter (mean diameter for an alloy) and the bulk $T_m$ as input parameters, we have also estimated the strongest sizes for some samples at room temperature. As listed in Table I, the predicted critical sizes are within 8~35 nm, agreeing exceedingly well with documented results.

3.4  Strongest size: dominating factors

Figure 4 plots the strongest size $K(x^{-2}_C)$ dependence on the factors of $A_0$, m and $T/T_m(0)$. The plots reveal the following:

(i)  The measured strongest size is not a constant but varies sensitively to the parameters of m and $T/T_m(0)$.

(ii) The $x_C$ depends less on the $A_0$ value if the $T/T_m(0)$ ratio is smaller than 0.2, which means that the critical size measured at very low temperature is

(iii) purely intrinsic and varies only with the bond nature. At relatively higher temperatures, extrinsic factors become non-negligible.

(iii) When the operation temperature is raised from zero to the $T_m(x)$, the strongest size drops from the maximum to a minimal and then turns up with the temperature in a "U" shape. Therefore, it is not surprising why the reported critical sizes for the same specimen vary from source to source because of the operating temperature difference. A slight change of the $T/T_m$ value may leads to substantial different sizes. At the same $T/T_m$ value, the materials with higher m values (or more covalent in bond nature) gives smaller critical sizes.

(iv) For a given critical size, there are at least two $T/T_m(0)$ values. However, the mechanical strength of the same critical size obtained at different temperatures is completely different, as illustrated in Figure 3b.

The predicted m, $A_0$, and $T/T_m(0)$ dependence of the $x_C$, $T_C(x)$, and the temperature trends of mechanical strength and compressibility/extensibility coincide exceedingly well with the cases as reported by Eskin et al[46] on the grain size dependence of the tensile elongation (extensibility) of an $Al_{0.04}Cu$ alloy in the quasisolid state. The ductility increases exponentially with temperature until infinity at $T_m$ that drops with solid size. On the other hand, the ductility increases generally with grain refinement. This is also the frequently observed cases such as the size-enhanced compressibility of alumina[47] and PbS[48] in nanometer range at room temperature.[49,50] The predicted trends also agree with experimental observations[51] of the temperature dependence of the yield stress of Mg nanosolids of a given size showing that the yield strength drops as the operating temperature is raised.

3.5    Correlation between elasticity and hardness: shape dependence

Theoretically, the analytical expressions for elasticity, stress, and hardness, should be identical in nature and in the same unit. However, the extrinsic competition between activation and resistance to glide dislocations in the plastic deformation differ the hardness from the elastic modulus or residual stress substantially. Such competition is absent from the elastic deformation in particular using the non-contact measurement techniques. Recent measurement[52] of the size dependence of the hardness, shear stress, and elastic modulus of copper nanoparticles, as show in Figure 6a, confirmed this expectation. The shear stress and elastic modulus of Cu reduce monotonically with the solid size but the hardness shows the strong IHPR at ~40 nm.

On the other hand, the IHPR effect is more pronounced to the harness of nanostructures than to the hardness-depth (P-d) profile of thin films. As compared in Figure 6b and c, the hardness and Young's modulus of Ni films are almost linearly interdependence.[53] The P-d profile deviates from the relationship of $P/P_0 = (1+K_0/K)^{0.5}$ proposed by Nix and Gao[54] for Vickers hardness with $K_0$ being an adjustable parameter. Strikingly, the surface hardening evidences for the broken bond induced energy density gain in the skin, which coincides with the IHPR though the surface passivation may present. Therefore, the competition between the inhibition and the activation of atomic dislocations triggers the IHPR of nanograins yet in the nanoindentation test of thin films the inhibition of dislocations my be dominance.

One may ague that, in the nanoindentation test, errors may arise because of the shapes and sizes of the tips. In practice, the stress-strain profiles of a nanosolid are not symmetrical when comparing the situation under tension to the situation under compression,[55] and the flow stress is dependent on strain rate, loading mode and time, and materials compactness, as well as size distribution. By taking the relative change of the measured quantity, artifacts because of the measuring techniques can be minimized in the present approach, seeking for the change relative to the bulk counterpart measured under the same conditions.

IV Summary

An analytical solution has been developed for the plastic yield strength of nanograins based on the combination of the BOLS correlation, Bern's criterion, and the conventional HPR and on the following physical constraints: (i) mechanical enhancement happens at the site surrounding a defect or at a surface because of the broken bonds induced local strain and trapping; (ii) melting point suppression at sites near the defects because of the atomic cohesive energy loss; (iii) atomic dislocation requires activation energy that is proportional to the melting point. Derived solution has enabled us to reproduce the observed HPR and IHPR and identify the factors dominating the strongest sizes. Conclusions can be drawn as follows:

(i) The IHPR originates from the broken bond induced lattice strain, energy density gain and the cohesive energy loss of nanograins.

(ii) The IHPR is dominated by the competition between the activation and the inhibition of atomic dislocations. The energy-density gain in the surface skin and the effect of strain work hardening are responsible for the inhibition of atomic dislocations yet the cohesive-energy loss caused by the under-coordinated GB atoms dominates the activation energy for the dislocations.

(iii) When the grain is greater than the strongest size, the process of dislocation inhibition is dominant; at the strongest size, the processes of activation and inhibition of dislocations contribute competitively; during the softening, contribution from the cohesive energy loss becomes dominant.

(iv) The self-heating during detection and the size induced presence of the soft quasisolid phase are responsible for the softening in the IHPR and for the superplasticity of a metallic nanosolid.

(v) The IHPR critical size is predictable. The critical size is dominated intrinsically by the bond nature indicator, the $T/T_m$ ratio, and extrinsically by experimental conditions or other factors such as size distribution and impurities that are represented by the factor $A_0$.

(vi) The IHPR at larger solid size converges to the normal HPR that holds its conventional meaning of the accumulation of atomic dislocations that resist further atomic displacements in plastic deformation. The slope in the traditional HPR is suggested be proportional to $\exp(T_m/T)$, which addresses the relationship between the hardness and the activation energy for atomic dislocations. The $K_j^{-0.5}$ in the conventional HPR should represent the accumulation of atomic dislocations that resists further dislocations.

The project is supported by NNSF of China (Nos. 10772157 and 10525211) and Ministry of Education (RG14/06), Singapore.

Table and Figure captions:

Table 1

Prediction of the critical sizes for IHPR transition with $A_0 = 0.5$ and $m = 1$ otherwise as indicated.

| Element | Measured $D_C$/nm | Predicted $D_C$/nm ($A_0 = 0.5$, $m = 1$) |
|---|---|---|
| Fe | 18.2 | 19.1 |
| Ni | 17.5 | 18.8 |
| Cu | 14.9 | 16.1 |
| Zn | 17.2 | 18.5 |
| Pd | 19.9 | 20.8 |
| C |  | 20.5 ($m = 2.56$) |
| Si | 9.1 | 10.6 ($m = 4.88$, $A_0 = 0.1$) |
| Ge |  | 11.5 ($m = 4.88$, $A_0 = 0.1$) |
| Al |  | 16.3 |
| Sn |  | 47.2 |
| Pb |  | 28.0 |
| Bi |  | 36.0 |
| $Ni_{80}P_{20}$ | 7.9 | 8.9 (m = 4) |
| $NiZr_2$ | 17.0 | 19.8 (m = 4) |
| $TiO_2$ | 22.5 | 23.1 (m = 4, $A_0$ = 0.01) |

Figure 1    The core-shell structure of a nanograin of K radius. The broken bonds induced surface strain and trapping causes localized densification of charge, energy and mass. The strained and stiffened skin provides pinning centre for inhibiting atomic dislocations at low temperatures. On the other hand, the bond order deficiency lowers the cohesive energy of atoms in the skin and hence the activation of dislocations. The competition between the energy density gain and the atomic cohesive energy loss originates intrinsically the IHPR that can only be initiated by the contact mode of plastic deformation detection.

Figure 2    Agreement between predictions and the measured size dependence of (a) energy density (elastic modulus) gain and (b) atomic cohesive energy loss ($T_m$ depression of Al and Sn thin plates melting and Au and Ag dot evaporating) of nanostructures.

Figure 3    (a) Size induced presence of the quasisolid phase and the corresponding critical temperatures. The $T_C$ for solid-quasisolid transition is much lower than the $T_m$ that was derived from the BOLS correlation and Bern's criterion for melting. The presence of the soft quasisolid phase and the self-heating during processing are suggested to be responsible for the size induced softening and superplasticity of nanostructures. (b) Temperature dependence of hardness of different critical sizes of the same material (m value).

Figure 4   Reproduction of the measured IHPR (scattered data) for (a) Ni [42] (b) TiO$_2$ [44] and NiP [43] in comparison to the classical HPR (straight lines) and the IHPR (broken lines) without $\varphi(d,m,x,T)$ contribution. The slope $A_0 = 0.5$ was optimised otherwise as indicated for all the samples.

Figure 5   (a) The slope $A_0$, (b) bond nature, and $T/T_m(0)$ dependence of the strongest size $K_C$. Minimal size is available at $T/T_m(0) \sim 0.2$-$0.4$. For a given $K_C$ there are at least two $T/T_m(0)$ values.

Figure 6 (a)  Comparison of the measured size dependence of the elasticity, shear stress and the hardness of Cu nanostructures [48], and the nanoindentation depth dependence of (b) the hardness and (c) the correlation between the hardness and elastic modulus of Ni films with 0 and 10% tensile strains[49], indicating the activation of the IHPR in the contact mode the plastic deformation detection of nanostructures and the dominance of inhibition in indentation depth profiling.

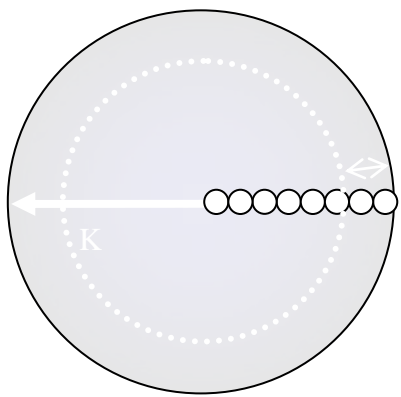

Fig 1

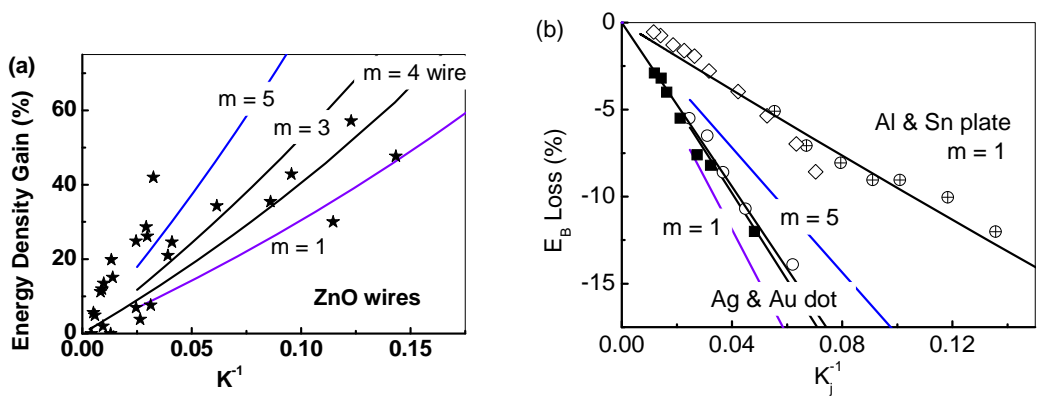

Fig 2 a, b

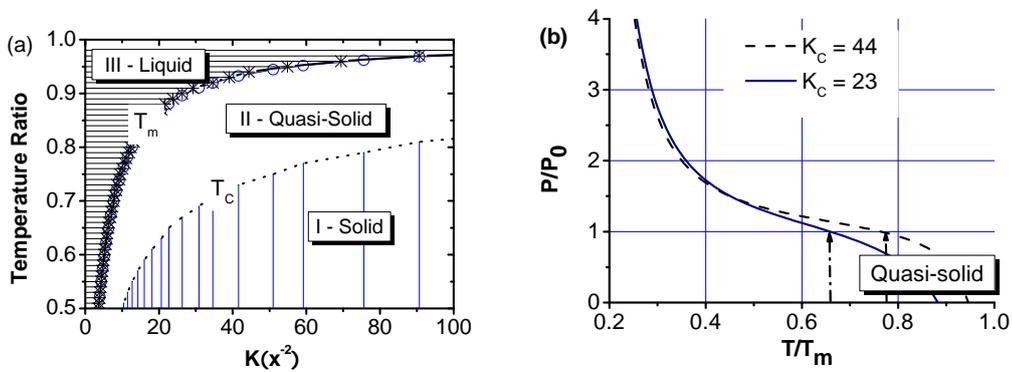

Fig 3

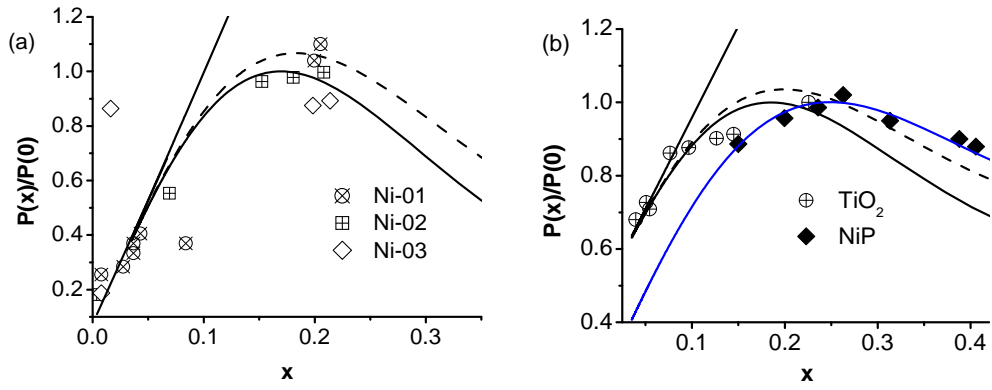

Fig 4 a, b

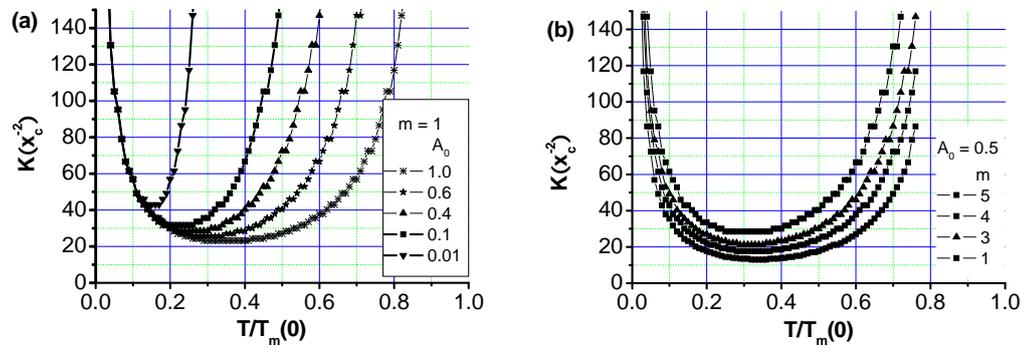

Fig 5 a, b

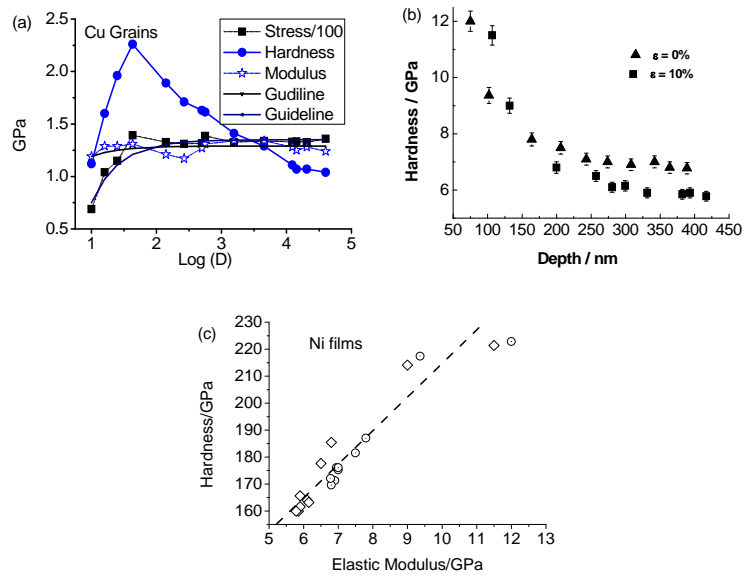

Fig 6 a-c